\documentstyle[12pt]{article}
\def\thesection{\arabic{section}}

\def\appendix{\setcounter{section}{0}
 \def\thesection{Appendix \Alph{section}}
 \def\theequation{\Alph{section}.\arabic{equation}}}
\newcommand{\beq}{\begin{equation}}
\newcommand{\eeq}{\end{equation}}
\newcommand{\bea}{\begin{eqnarray}}

\begin{document}
\begin{flushright}
Imperial/TP/94-95/42
\end{flushright}
\begin{center}
\vspace{1.0cm}
{\Large\bf Analytic properties of the QCD running  \\
\vspace{0.3cm}
coupling constant and $\tau$ decay} \\
\vspace{1.5cm}
{\large\bf     H.F. Jones} \\
\vspace{0.2cm}
{\em Physics Department, Imperial College, London SW7 2BZ,\\
United Kingdom }

\vspace{0.2cm}
{\large{\bf I.L. Solovtsov  and O.P. Solovtsova} }\\
\vspace{0.2cm}
{\em  Bogoliubov Laboratory  of  Theoretical  Physics, \\
Joint  Institute  for Nuclear Research, \\
Dubna, Moscow region, 141980, Russia } \\
\end{center}
\vspace{2.0cm}
\begin{abstract}
A non-perturbative expansion method which gives a well-defined
analytic continuation of the running coupling constant from
the spacelike to the timelike region is applied to the inclusive
semileptonic decay of the $\tau$--lepton. The method allows us to
evaluate $R_{\tau}$ by integration over the non-perturbative physical
region, rather than by using Cauchy's theorem, and hence to incorporate
threshold effects in a very direct way. Within our
framework the difference between the effective coupling constants in the
timelike and spacelike domains can be substantial and is not simply a matter
of the standard $\pi^2$ terms.
\end{abstract}

\newpage

The $\tau$ decay process with hadronic final states represents an important
test of quantum chromodynamics.  Due to the inclusive character of the
process, the ratio
\begin{equation}
\label{Eq.1}
R_{\tau}\, =\, \frac{\Gamma (\,\tau^- \to \nu_{\tau}\, {\rm hadrons}
 (\gamma)\,)}{
\Gamma (\,\tau^- \to \nu_{\tau}\, e^-\,\overline{\nu}_{e}\, (\gamma)\,)}\, \, ,
\end{equation}
where ($\gamma$) denotes possible additional photons or lepton pairs, is a
very convenient quantity
both for a theoretical investigation and for the definition of the QCD
coupling constant $\alpha_{s}(M_{\tau}^2)$.
A detailed theoretical analysis of this
problem has been given in Ref. [1] (~see also Refs.~[2-5], in which
different aspects of the problem are discussed~).

The starting point of the theoretical analysis is the expression
\begin{equation}
\label{Eq.2}
R_{\tau}  =  2 \int_{0}^{M^{2}_{\tau}} \, \frac{ds}{M^{2}_{\tau}} \,
\left(1 - \frac{s}{M^{2}_{\tau}}\right)^{2} \,
\left(1 + \frac{2s}{M^{2}_{\tau}}\right)\,
\tilde{R}(s)  \,\, ,
\end{equation}
where
\begin{eqnarray}
\label{Eq.3}
\tilde{R}(s)\,&=&\,\frac{\rm {N}}{2\pi\rm {i}}\,
\bigl[\,  \Pi(s+{\rm i}\epsilon)\,-\,
\Pi(s-{\rm i}\epsilon)\,\bigr] \,\, ,  \nonumber \\
\Pi (s)\,&=&\,\sum_{q=d,s}{\vert V_{uq} \vert}^2
\bigl ( \, \Pi_{uq,V}(s)\,+\,\Pi_{uq,A}(s) \,\bigr ) \,\, .
\end{eqnarray}
The normalization factor N is defined so that in zeroth order
perturbation theory $\tilde R^{(0)}_{\rm pert}=3$.
In the framework of standard perturbation theory the integral (2)
cannot be evaluated directly since the integration region in (2) includes
small values of momentum for which perturbation theory is
invalid\footnote{  In Ref.[6], the integral (2) has been calculated within
the method of optimized perturbative series [7].}.
Instead of Eq. (2), the expression for $R_{\tau}$ may be rewritten,
using Cauchy's theorem, as a contour integral in the complex $s$-plane
with the contour running clockwise around the circle
$\vert s\vert= M_{\tau}^2$. It seems that this trick allows
one to avoid the problem of calculating the  nonperturbative contribution,
which is needed if one uses Eq. (2).
However, the application of Cauchy's theorem is based on specific
analytic properties of $\Pi$(s) or the Adler $D$ function
\begin{equation}
\label{Eq.4}
D(q^2)\,=\,q^2\,(-\,\frac{d}{d\,q^2})\,{\rm {N}}\,\Pi(q^2)  \, .
\end{equation}
The function $D(q^2)$ is
an analytic function in the complex $q^2$--plane with a cut along
the positive real axis. It is clear that the
approximation of the $D$--function by perturbation theory breaks these
analytic properties. For example, the one-loop approximation for the QCD
running coupling constant has a singularity at $Q^2=\Lambda^2_{QCD}$,
the existence of which prevents the application of Cauchy's theorem.
Moreover, to define the
running coupling constant in the timelike domain,
one usually uses the dispersion
relation for the $D$ function derived on the basis of the
above-mentioned analytic properties. In the framework of perturbation theory,
this method gives the so-called $\,\pi^2$--term contribution
which plays an important role in the analysis of various processes~[8-12].
However, the same problem arises: the perturbative approximation
breaks the analytic properties of $\lambda^{\rm eff}(q^2)$
which are required to write the dispersion relation.
In addition, there is the problem of taking
account of threshold effects. As follows from Eq.~(2), the initial
expression for $R_{\tau}$ ``knows'' about the threshold; for example, the
value $R_{\tau}$ written in the form of Eq.~(2) contains the information
that the CKM mixing matrix element $V_{us}\,=\,0$ for $s\,<\,(m_u+m_s)^2$.
But all the threshold information is lost if one rewrites Eq.~(2)
as a contour integral and uses a fixed number of flavours for the calculation
of $\tilde R(s)$ on this contour.

In this note we will concentrate on both aspects of the problem.
Our considerations are based on a method of construction of
the effective running coupling constant which respects the
above-mentioned analytic properties [13,14].
In Ref.~[15] we have demonstrated that in the framework of this approach
there exists a well-defined procedure for defining the running
coupling in the timelike domain which does not conflict with the
dispersion relation. We will use the following definitions:
$\lambda^{\rm eff} =\alpha_{\rm  QCD}/(4 \pi ) \, $
is the initial effective coupling constant in the $t$--channel
(~spacelike region~) and $\lambda^{\rm eff}_s$
is the effective coupling constant in the $s$--channel (~timelike region~).
{}From the dispersion relation for the $D$--function we obtain~[15]
\begin{equation}
\label{Eq.5}
\lambda^{\rm eff}(q^2)\,=\,-\,q^2\,\int_0^{\infty}\,
\frac{d\,s}{{(s-q^2)}^2}\,\lambda_s^{\rm eff}(s)\,\, .
\end{equation}

Thus, the initial running coupling constant $\lambda^{\rm eff}(q^2)$ is
an analytic function in the complex $q^2$--plane with a cut along the
positive real axis. This function does not exist for real positive $q^2$,
so the definition of the running coupling constant in the
timelike domain is a crucial question. Here we use the standard
definition of $\lambda_s^{\rm eff}(s)$ in the $s$--channel based on
the dispersion relation for the Adler $D$--function. In this case,
parametrization of timelike quantities, for example
$R_{e^+e^-}(s)$ or $\tilde R(s)$, by the function
$\lambda_s^{\rm eff}(s)$
is similar to parametrization of spacelike processes by the
function $\lambda^{\rm eff}(q^2)$.

The inverse relation of Eq.~(5), given the analytic properties
of $\lambda^{\rm eff}(q^2)$, is of the form
\begin{equation}
\label{Eq.6}
\lambda_s^{\rm eff}(s)\,=\,-\,\frac{1}{2\pi {\rm i}}\,
\int _{s-{\rm i}\,\epsilon} ^{s+{\rm i}\,\epsilon} \frac{d\,q^2}{q^2}\,
\lambda^{\rm eff}(q^2)  \,\, ,
\end{equation}
where the contour goes from the point $q^2 =s-{\rm i}\epsilon$ to the point
$q^2 =s+{\rm i}\epsilon$ and lies in the region where
$\lambda^{\rm eff}(q^2) $ is an analytic function of $q^2$.
Equation~(6) defines the running coupling constant in the timelike
region which we must use to calculate $\tilde R(s)$ in Eq.~(2).
As follows from Eqs.~(5) and (6), there is a connection between the
asymptotic behaviours of $\lambda^{\rm eff}(q^2)$ and
$\lambda_s^{\rm eff}(s)$. If the function $\lambda^{\rm eff}(q^2) $
has the asymptotic behaviour
$$\lambda^{\rm eff}(q^2) \, \sim \, \frac{1}{b_0\,\ln(-q^2/\Lambda^2)}
\qquad {\rm as} \qquad q^2\,\ss\, -\, \infty$$
the function $\lambda_s^{\rm eff}(s)$ behaves like
$$\lambda_s^{\rm eff}(s) \, \sim \, \frac{1}{b_0\,\ln(s/\Lambda^2)}
\qquad {\rm as} \qquad s\,\ss\, +\, \infty\,\, $$
and vice versa.

To write Eq.~(6), it was important that the function
$\lambda^{\rm eff}(q^2)$ had the above-mentioned analytic properties.
For example, to use the one-loop approximation, one needs to
modify its infrared behaviour at $Q^2=\Lambda^2$ in an {\it ad hoc}
manner so that the singularity
at $Q^2=\Lambda^2$ is absent in the new expression for $\lambda (Q^2)$ .
A self-consistent formulation of the analytic continuation problem is,
however, possible within the scope of a systematic non-perturbative approach.

Let us rewrite Eq.~(2) in the form
\begin{equation}
\label{Eq.7}
R_{\tau}\,=\,2\int_0^1\,dx\, {(1\,-\,x)}^2\,(1+2\,x)\,
\tilde R(M_{\tau}^2\,x)\,\, ,
\end{equation}
where $\tilde R(M_{\tau}^2\,x)\,$ can be expanded as the series
\begin{eqnarray}
\label{Eq.8}
\tilde R(M_{\tau}^2\,x)\,&=&\,r_0\,\bigl[\,1\,+\,
r_1\,\lambda_s(M_{\tau}^2\,x)\,+\,r_2\,\lambda_s^2(M_{\tau}^2\,x)\,
+\,\cdots \bigr]  \nonumber\\
&=&\,r_0\,\bigl[\,1\,+\,r_1\,\lambda_s^{\rm eff}(M_{\tau}^2\,x)\,\bigr]\, .
\end{eqnarray}

Here we have introduced the effective coupling constant in the
$s$--channel, which incorporates higher-order corrections. As has been
mentioned above, any approximation of the quantity  $\tilde R$
by a finite number of terms of the perturbative series does not allow us
to calculate the integral (7) directly because this expression includes
the integration region of small $s$ where perturbation theory
is invalid.
In Ref.~[15], the $s$--channel effective coupling constant has been
constructed within the nonperturbative method
proposed\footnote{  In Ref.[16] the method has been applied to
describe the process of $e^+e^-$ annihilation at low energies.}
in Refs.~[13,14]. This approach to QCD allows one
to systematically investigate the low energy structure of the theory
and evaluate the integral (7) directly without the use of Cauchy's theorem.
The method is based on a new expansion parameter $a$ which is connected
with the original coupling constant by the following equation
\begin{equation}
\label{Eq.9}
\lambda\,=\,\frac{1}{C}\,\frac{a^2}{(1\,-\,a)^3}\,\, ,
\end{equation}
where $C$ is a positive parameter which we will fix on the
basis of meson spectroscopy [14]. From Eq.~(8) it is easy to see
that the expansion parameter $a$ is bounded by
$0\,\leq\,a\, <\,1$ for all values of the
initial coupling constant in $0 \leq\lambda < \infty $.

The $Q^2$--evolution of the parameter $a$ is defined by
\begin{equation}
\label{Eq.10}
f(a)\,=\,f(a_0)\,+\,\frac{2\,b_0}{C}\,\ln\,\frac{Q^2}{Q_0^2}\,\, ,
\end{equation}
where $a_0$ is the value of the parameter $a$ at some normalization
point $Q_0$ and to order $a^3$ the function $f(a)$ has the form
\begin{equation}
\label{Eq.11}
f(a)\,=\,\frac{2}{a^2}- \frac{6}{a} - 48\ln a-
\frac{18}{11}\,\frac{1}{1-a} +  \frac{624}{121}\,\ln{(1-a) } +
\frac{5184}{121}\,\ln{( 1+\frac{9}{2}\,a )} \, .
\end{equation}

The effective coupling constant in the $s$--channel can be written
as
\begin{equation}
\label{Eq.12}
\lambda_s(s)\,=\,\frac{1}{2\pi\,b_0}\,{\rm Im}\,{\phi}(a_{+})\, \, ,
\end{equation}
where
\begin{equation}
\label{Eq.13}
{\phi}(a)\,=\,-4\,\ln \,a\,-\,\frac{72}{11}\,\frac{1}{1\,-\,a}\,+\,
\frac{318}{121}\, \ln\,(1\,-\,a)\,+\,\frac{256}{363}\,
\ln \,(1\,+\,\frac{9}{2}\,a)\,\,  ,
\end{equation}
and the value of $a_{+}$ obeys the following equation:
\begin{equation}
\label{Eq.14}
f(a_{+})\,=\,f(a_0)\,+\,\frac{2\,b_0}{C}\,
\bigl[\,\ln\, \frac{s}{Q_0^2}\,+\,{\rm i} \,\pi\,\bigr]\,\, .
\end{equation}

In the calculation of the integral (7)
the number of active quarks is different in various regions of the
integration. One usually applies the matching procedure in the Euclidean
$t$--channel, changing the number of active quarks at the threshold
$Q\,=\,\xi\,m_q$ with a matching parameter
$1\,\leq\,\xi\,\leq\,2$ [17] (~see also Refs.~[18,19]~).
However, within the framework of a massless renormalization scheme, a
more natural way is to introduce the threshold matching in the physical
$s$--channel, where, at least in leading order, the change in
the number of active quarks is an obvious fact associated
with the energy threshold of the quark pair production.
Moreover, any matching procedure for the coupling constant in the
$t$--channel, for which one uses a condition of the type
$ {\rm Re}\,Q^2 > \xi^2\,m_q^2 $,
leads to a violation of the analytic properties of $\lambda^{\rm eff}(q^2)$,
whereas the matching procedure in the $s$--channel maintains the
analytic properties of the effective coupling constant
$\lambda^{\rm eff}(q^2)$ in the complex $q^2$--plane.

By using the method under consideration we can ensure that both
$\lambda_s(s)$ and its derivative $\lambda_s'(s)$ are
continuous at various threshold points $\tilde s_i$.
The set of corresponding equations for the parameters
$C^{(f)}$ and $a_0^{(f)}$ is as follows:
\begin{eqnarray}
\label{Eq.15}
&&\frac{1}{b_0(f-1)}\,{\rm Im}\,\phi (a_+^{(f-1)})\,=\,
\frac{1}{b_0(f)}\,{\rm Im}\,\phi (a_+^{(f)})\,\, ,            \\
&&\frac{1}{C^{(f-1)}}\,{\rm Im}\,\bigl[\,
{\bigl(a_+^{(f-1)}\bigr)}^2 \,\bigl(1\,+\,3\,a_+^{(f-1)}\bigr)\,\bigr]\,
=\,
\frac{1}{C^{(f)}}\,{\rm Im}\,\bigl[\,
{\bigl(a_+^{(f)}\bigr)}^2 \,\bigl(1\,+\,3\,a_+^{(f)}\bigr)\,\bigr]\,
\nonumber
\end{eqnarray}

For the number of active flavours $f\,=\,3$ we use the value
$C^{(3)}\,=\,4.1$  obtained from the phenomenology of
meson spectroscopy [14]. Then for $f\,\ne\,3$ we find the parameter
$C^{(f)}$ using Eqs.~(15). We use for the matching parameter
$\xi$ the value $\xi\,=\,2$, but the value of $R_{\tau}$ from Eq.~(7) is
practically independent\footnote{ The corresponding uncertainty is smaller
than 0.01\% .}
of the parameter $\xi$ in the interval $1\leq \xi \leq 2$.

 In the alternative approach to the evaluation of $R_\tau$ using
Cauchy's theorem one proceeds first by an integration by parts to
convert $\tilde R$ into $D$, then represents the discontinuity
as a contour integral and finally opens up the contour to the
unit circle in the $z$ plane. In this way $R_\tau$ is expressed
in terms of $\lambda^{\rm eff}$ as
\begin{equation}
\label{Eq.16}
R_{\tau}\,=\,\frac{1}{2\pi {\rm i}}\,
\oint_{|z|=1}\,\frac{dz}{z}\,{(1\,-\,z)}^3\,(1\,+\,z)\,
D(M^2_{\tau}z)\, \, ,
\end{equation}
where the $D$ function has the following form
\begin{equation}
\label{Eq.17}
D(M^2_{\tau}z)\,=\,d_0\,\bigl[\,1\,+\,d_1\,\lambda^{\rm eff}(M^2_{\tau}z)\,
\bigr]\, ,
\end{equation}
and to order $O(a^3)$
\footnote{ In $O(a^2)$ we obtain similar results, in accordance with
the mechanism of induced convergence (for a more detailed account
see Ref.~[15]).}
\begin{equation}
\label{Eq.18}
\lambda^{\rm eff}(Q^2)\,=\,\frac{1}{C}\,a^2\,(1\,+\,3a)\, ,
\end{equation}
where $a=a(Q^2)$ is found from Eq.~(10).
We can rewrite Eq.~(16) in the form
\begin{equation}
\label{Eq.19}
R_{\tau}\,=\,R_{\tau}^{(0)}\,(\,1+\,\Delta R_{\tau}\,)\,\, ,
\end{equation}
with $R_{\tau}^{(0)}$ defined as
\begin{equation}
\label{Eq.20}
R_{\tau}^{(0)}\,=\, 3\,(\,|V_{ud}|^{2}\,+\,|V_{us}|^{2}\,)\,S_{\rm {EW}} \, ,
\end{equation}
where the electroweak factor $S_{\rm {EW}}$ and the CKM matrix elements
$|V_{ud}|$, $|V_{us}|$ are
$  S_{\rm {EW}}\,=\,1.0194\; , \; |V_{ud}|\,=\,0.9753 \; ,
\;|V_{us}|=0.221 \; $  taken from Ref.~[1].

\noindent
For $\Delta R_{\tau}$ one finds
\begin{equation}
\label{Eq.21}
\Delta R_{\tau}\,=\,\frac{1}{2\pi {\rm i}}\,d_1\,
\oint_{|z|=1}\,\frac{dz}{z}\,{(1-z)}^3\,(1+z)\,
\lambda^{\rm eff}\,(M_{\tau}^2\,z)\, .
\end{equation}
By using the parametrization $z\,=\,-\,M_{\tau}^2\, e^{i\theta}$
and with $d_1=r_1=4$ we get
\begin{equation}
\label{Eq.22}
\Delta R_{\tau}=\frac{2}{\pi}\int_{-\pi}^{\pi}d\theta
\bigl ( 1+2e^{i\theta}-2e^{3i\theta}-e^{4i\theta} \bigr )\,
\lambda^{\rm eff} \bigl ( M_{\tau}^{2}e^{i\theta} \bigl ) \,.
\end{equation}

It should be stressed that we can use Eq.~(22) only for a fixed
number of quark flavours. Indeed, we have applied the physical matching
procedure in the $s$--channel and have obtained an analytic function
$\lambda^{\rm eff}\,(q^2)$ with a cut along the positive
real axis. This function ``knows", in principle, about all quark
thresholds due to Eq.~(5). For a function $\lambda^{\rm eff}\,(q^2)$
obtained in such a way it is impossible to split the region of
$Q^2\,=\,-\,q^2$ into subregions in which the number of active quarks
is fixed. The situation is similar to the situation that arises
in the mass-dependent momentum renormalization scheme. In this
scheme, all quarks contribute to the effective coupling
constant; however, for $Q^2\,<<\, M_q^2$ the quark with mass $M_q$
is irrelevant because its contribution to the running coupling is very small.

To check the consistency of our method let us first consider a fixed number
of active quarks
$f=3$  and use  $R_{\tau}=3.56$ as an input. In this case,
using Eqs.~(7),(8),(12)-(14)
we can find
the parameter $a_0$ and then verify that Eqs.~(7) and (16)
give the same results for $R_{\tau}$: $R_{\tau}=3.560$ as should be
the case.

To take into account the threshold effects, we have used
Eqs.~(15) to find the parameter $C^{(f)}$ for $f\ne 3$ and the conditions
that the CKM matrix elements $V_{ud}=0$ for $s < (m_u+m_d)^2$ and
$V_{us}=0$ for $s < (m_u+m_s)^2$.
We use here the following values of the quark masses:
$m_u=5.6\, \rm {MeV}$, $m_d=9.9\, \rm {MeV}$ and  $m_s=199\, \rm {MeV}$.
In this way, we obtain from Eq.~(7) for $R_{\tau}$ the value 3.552 instead
of 3.560. One can see that the
threshold effects for $R_{\tau}$ are about $0.2\; \%$.

Now, taking the experimental value $R_{\tau}=3.552$ [20] as an input,
we  obtain
$\alpha_s(M^2_{\tau})\,=\,0.37$ and  $\alpha(M^2_{\tau})\,=\,0.40$.
The values of the coupling constant in the $s$-
and $t$--channels are clearly different from each other;
the ratio is  $\alpha_s(M^2_{\tau}) / \alpha(M^2_{\tau})=0.92$.

The experimentally measurable quantity $R_{\tau}$ can be parametrized both
by the function $\alpha_s(s)$ defined in the time--like region and
entering into the initial expression for $R_{\tau}$ (Eq.~(2)) and by the
running coupling constant $\alpha(q^2)$ used in the contour integral
(Eq.~(16)). The perturbative expansion does not allow one to perform the
integration in Eq.(2) directly because it involves a non-perturbative
region. Instead, one usually uses the perturbative formula to evaluate
the contour integral (16). However, we believe this to be inconsistent
because the analytic properties which are required to write down the
Cauchy integral are not respected by the perturbative formula.

In the present paper we have proposed a method which allows one to
evaluate both the initial integral for $R_{\tau}$ and the expression
obtained by the use of Cauchy's theorem. They are equal in the absence
of threshold effects, which, however, can be easily incorporated in
the first method. We have also demonstrated that the distinction
between the functions $\alpha_s(s)$ and $\alpha(q^2)$ is not simply a
matter of the standard $\pi^2$ terms, which may be important for
understanding certain discrepancies [19, 21] arising in the
determination of the QCD coupling constant from various experiments.
For a more detailed quantitative analysis of the experimental data using
our method it will be necessary to include higher-order corrections.
In the near future we plan to extend the calculation to O($a^5$).


\vspace{0.5cm}
The authors would like to thank D.I.~Kazakov and A.N.~Sissakian for
interest in the work and useful comments.

\end{document}